\documentclass[apj]{emulateapj-rtx4}

\usepackage{amssymb}
\usepackage{soul}
\usepackage{graphicx}
\usepackage{amsmath}
\usepackage[usenames]{color} 
\usepackage{ulem}
\usepackage{lscape} 
\usepackage{rotating}
\def\gv{2002\,GV\ensuremath{_{31}}}
\def\wg{2010\,WG\ensuremath{_{9}}}
\def\jj{(278361)\,2007\,JJ\ensuremath{_{43}}}
\def\jjs{2007\,JJ\ensuremath{_{43}}}

\begin{document}
\sloppy

\title{Pushing the limits: K2 observations of the trans-Neptunian objects \gv{} and \jj{}}
\shorttitle{TNO observations with K2}

\author{A.~P\'al\altaffilmark{1,2}}
\email{apal@szofi.net}
\author{R.~Szab\'o\altaffilmark{1}}
\author{Gy.~M.~Szab\'o\altaffilmark{1,3,4}}
\author{L.~L.~Kiss\altaffilmark{1,4,5}}
\author{L.~Moln\'ar\altaffilmark{1}}
\author{K.~S\'arneczky\altaffilmark{1,4}}
\author{Cs.~Kiss\altaffilmark{1}}
\altaffiltext{1}{Konkoly Observatory, Research Centre for Astronomy and Earth Sciences, Hungarian Academy of Sciences, H-1121 Budapest, Konkoly Thege Mikl\'os \'ut 15-17, Hungary}
\altaffiltext{2}{E\"otv\"os Lor\'and Tudom\'anyegyetem, H-1117 P\'azm\'any P\'eter s\'et\'any 1/A, Budapest, Hungary}
\altaffiltext{3}{ELTE Gothard Astrophysical Observatory, H-9704 Szombathely, Szent Imre herceg \'ut 112, Hungary}
\altaffiltext{4}{Gothard-Lend\"ulet Research Team, H-9704 Szombathely, Szent Imre herceg \'ut 112, Hungary}
\altaffiltext{5}{Sydney Institute for Astronomy, School of Physics A28, University of Sydney, NSW 2006, Australia}

\begin{abstract}
We present the first photometric observations of trans-Neptunian objects 
(TNOs) taken with the {\it Kepler} space telescope, obtained in the course 
of the K2 ecliptic survey. Two faint objects have been monitored in 
specifically designed pixel masks that were centered on the stationary 
points of the objects, when their daily motion was the slowest. In the 
design of the experiment, only the apparent path of these objects were 
retrieved from the detectors, i.e. the costs in terms of Kepler pixels 
were minimized. Because of the faintness of the targets we employ specific 
reduction techniques and co-added images. We measure rotational periods and 
amplitudes in the unfiltered {\it Kepler} band as follows: for \jj{} and 
\gv{} we get $P_{rot}= 12.097$\,h and $P_{rot}= 29.2$\,h while
$0.10$ and $0.35$ mag for the total amplitudes, respectively. 
Future space missions, like TESS and PLATO are not well suited to this 
kind of observations. Therefore, we encourage to include the brightest 
TNOs around their stationary points in each observing campaign to exploit 
this unique capability of the K2 Mission -- and therefore to provide unbiased 
rotational, shape and albedo characteristics of many objects. 
\end{abstract}

\keywords{	methods: observational --- 
		techniques: photometric --- 
		astrometry --- 
		minor planets, asteroids: general --- 
		Kuiper belt objects: individual  (\gv{}, \jj{}) }

\section{Introduction}
\label{sec:introduction}

\begin{figure*}
\centering\includegraphics[width=17.0cm,angle=0]{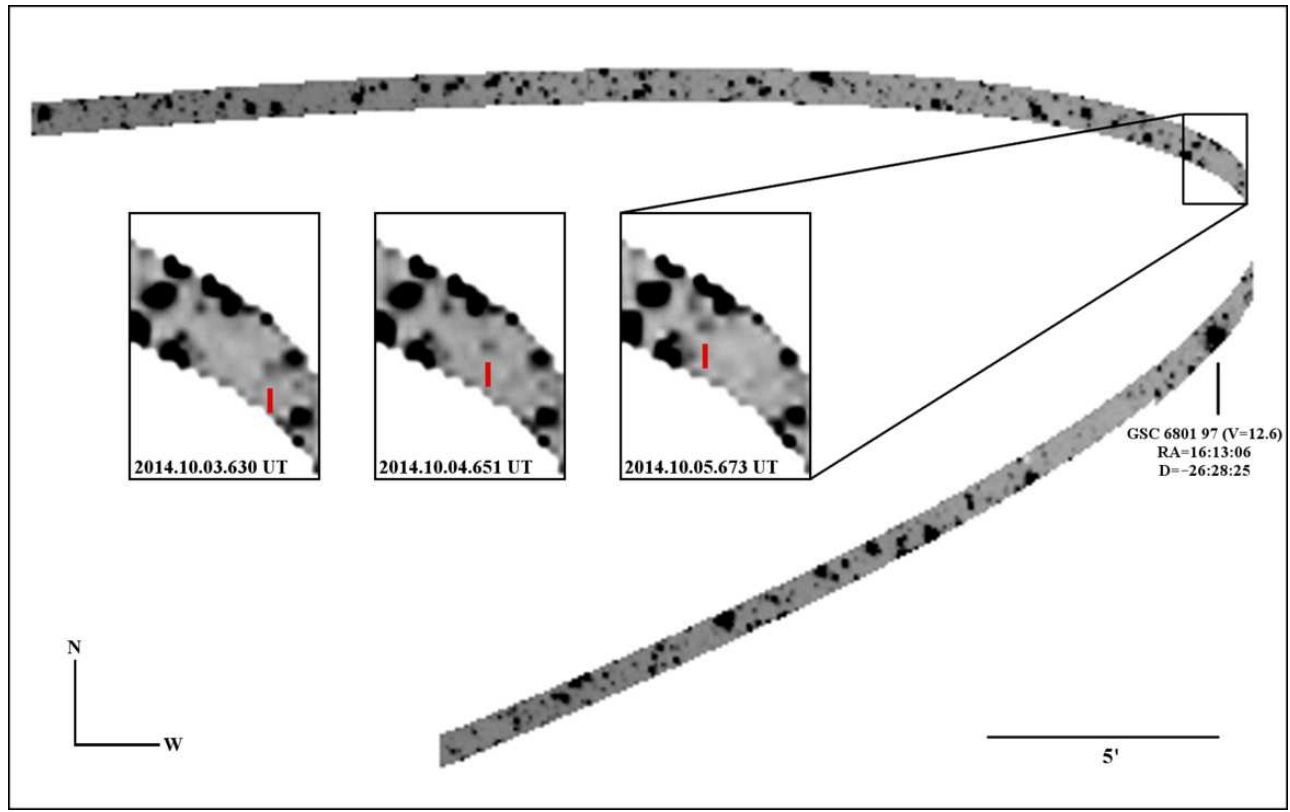}
\caption{Trajectory of \jj{} and the allocated 661 $1\times11$ and $1\times13$
pixel stripes on the CCD module \#17 of the {\it Kepler} space telescope. 
In the insets we show three individual exposures along the arc after 
the stationary point which unfortunately fell off silicon. The small inserts
show an area of $1.3^\prime\times1.8^\prime$.}
\label{fig:traj}
\end{figure*}

{\it Kepler} has provided incredible results on extrasolar planets and 
planetary systems, as well as on stellar astrophysics. After analyzing 
main-belt asteroids \citep{szabo2015}, here we continue the exploration 
of our own Solar System by pushing the limits of the spacecraft and 
observing faint, trans-Neptunian objects. 

The {\it Kepler} spacecraft is equipped with a 0.95-m Schmidt 
telescope and 42 imaging CCDs, from which 38 are currently in 
operation. {\it Kepler} was designed to be the most precise 
photometer in order to detect the transits of numerous small 
planets \citep{borucki2010}. It monitored approximately 
170~000 targets in the Lyra-Cygnus field, at high Ecliptic latitudes, 
almost continuously for four years during its primary mission. But 
after the failure of two reaction wheels the telescope permanently 
lost its  ability to maintain that attitude. Soon a new mission, 
called K2, was initiated to save the otherwise healthy and capable 
space telescope \citep{howell2014}. Since then, {\it Kepler} has 
been observing in shorter, 75-day-long campaigns along the Ecliptic 
to balance the radiation pressure from the Sun.

The ecliptic fields include innumerable Solar System objects, although 
most asteroids and planets cross the field-of-view quite quickly. TNOs, 
in contrast, appear to move slowly around their stationary points, and 
can be observed for extended periods of time. The $4\arcsec/{\rm pixel}$ 
resolution of the CCDs is sufficient to avoid confusion even for faint 
targets. But the observations posed various challenges for {\it Kepler}. 
Even the brighter TNOs are at the detection limit of the telescope 
($V=20-22\,{\rm mag}$), with an expected precision of a few tenths of a 
magnitude for a single long-cadence (30-min) observation. Moreover, TNOs 
exhibit quadratically increasing proper motions away from the stationary 
points. Since the pixel mask allocation is fixed for an entire campaign, 
the whole length of the orbital arc has to be observed continuously. 
Despite these challenges, the K2 mission is in a unique position to 
gather continuous TNO light curves in white light. 

The rotational light curve of small bodies in the Solar System is 
determined by the shape and albedo variegations on the surface of the 
target, and provides information on these important characteristics. In 
addition, the collisional history of these bodies is stored in the spin 
state of the objects, probing different eras of the Solar Systems's 
debris disk evolution depending on the size. The largest objects 
($\sim500\,{\rm km}$ or larger in diameter) decoupled early from 
collisional evolution of the debris disks, therefore their rotation 
should reflect the cumulative effect of collisions during the formation 
period. Medium size objects ($100-500\,{\rm km}$) likely avoided 
catastrophic impacts with a rotation affected by smaller past events, 
while the smallest bodies ($<100\,{\rm km}$) are probably remnants of 
several collisions during the evolution of the Solar System's debris 
disc \citep[for more details, see also][]{lacerda2005}.

In order to study the rotational properties of TNOs we proposed a pilot 
study for K2 Campaign~1 to observe the object \gv{} ($V\approx 
22.5\,{\rm mag}$) with a small pixel mask for a limited time around the 
stationary point (proposal ID: GO1064). The field of Campaign~2 included 
multiple moderately bright targets from which ultimately only \jj{} 
($V\approx 20.25$ mag) was selected for observation (Proposal IDs GO2066 
and GO3053).

\section{Observations}
\label{sec:observations}

For \gv{} a rectangular, $23 \times 22$-pixel mask was allocated by the 
Guest Observer Office that included the object for 16~days around 
its stationary point, but 2.8~days were lost to a mid-campaign data 
download period. In contrast, \jjs{} 
was covered with  a long arc constructed from a mosaic of 661 small 
($1\times11$ or $1\times13$) pixel masks, but the region of the stationary 
point itself fell off silicon. This caused a $20.9\,{\rm d}$
long gap in the observations, separating the data into a $18.7\,{\rm d}$ and a
$39.2\,{\rm d}$ long section.

The elongation of the object decreased from 
135 to 60 degrees during the campaign. The public target pixel time series 
files from the C1 and C2 fields were retrieved from the 
MAST archive\footnote{https://archive.stsci.edu/k2/} for the respective 
observations.

Without stabilization around the third axis, the {\it Kepler} space 
telescope slowly rolls about the optical axis. This movement causes the 
field-of-view to rotate slightly and it is corrected by the on-board 
thrusters at every $5.88\,{\rm h}$ ($f_{\rm corr} = 4.08\,{\rm d}^{-1}$). 
The spacecraft stores the pixels of pre-selected targets only, similarly
to the primary mission, but larger target pixel masks are allocated
for less targets in the K2 fields to accommodate the larger pointing
jitter and less frequent data download periods. Hence, the data acquisition
principles are exactly similar to the stellar targets, only the combination
of the small pixel masks of \jjs{} required additional steps during analysis. 

The attitude of the spacecraft was adjusted by a few pixels shortly after 
the start of both campaigns. This did not affect the data of \gv{} but 
in the first 49 frames of Campaign~2, \jjs{} was too close to the edge 
of the mask and those were discarded. Apart from these adjustments, the 
largest correction movements were always smaller than a pixel and the root mean squre 
of image shift offsets were $0.3$\,pixels for both campaigns. 
A closer inspection of the field-of-view movements 
revealed that the direction of motion reversed at the middle of the 
campaign and in some cases the correction maneuver did not occur and 
the telescope drifted for $11.76\,{\rm h}$ (instead of $5.88\,{\rm h}$).

\section{Data reduction}
\label{data}

\begin{figure}
\centering\includegraphics[width=8.5cm,angle=0]{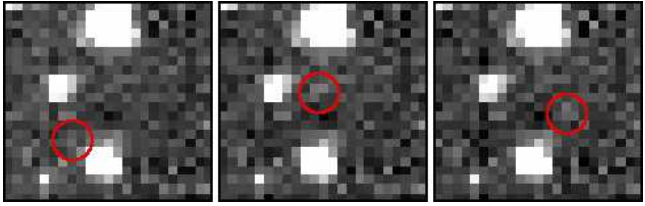}
\caption{Detection of the motion of the faint TNO, 
\gv{} around its stationary point, on 1-day co-added images. 
These stamps are centered at $\alpha=11^{\rm h}28^{\rm m}38.8^{\rm s}$, 
$\delta=+05^\circ34^\prime02^{\prime\prime}$ and cover the 
$23\times 22$ {\it Kepler} pixels, yielding a field-of-view of 
nearly $1.5^\prime\times1.5^\prime$. The image orientation is related to 
{\it Kepler} sensors during these observations, i.e. north is roughly 
to the left while east is downwards.}
\label{fig:gvimage}
\end{figure}

While in the case of \gv{}, the target pixel file contained a single stamp
of $23\times 22$ pixels, the data extraction was much complex in the 
case of \jjs{}. Here, the total mask related to this object has been splitted
to individual target stamps containing only $1\times 11$ or $1\times 13$ pixels.
Hence, the first step of the reduction was the reconstruction of the effective
section of the target CCD areas and the marking of the outlier pixels 
(in order to exclude them from the further data processing). 
In the case of \gv{}, we employed the PyKE package \citep{sb2012} to 
create individual FITS frames for each timestamp while the combined 
and marked frame series for \jjs{} was created using the utilities of 
the FITSH\footnote{http://fitsh.szofi.net/} package \citep{pal2012}.
Prior to the light curve analysis (see Sec.~\ref{sec:analysis}), for the subsequent 
data processing steps  we employed various tasks of the FITSH package 
(esp. \texttt{fiign}, \texttt{fitrans}, \texttt{grtrans}, \texttt{grmatch}, 
\texttt{fiphot}, \texttt{fistar}).

Since the target pixel time series do not contain information regarding
to the time variation of the astrometric solutions of the individual smaller
stamps, frame registration had to be based on purely imaging data. 
Due to the fairly strange shape of the field containing the apparent path
of \jjs{} (see. Fig.~\ref{fig:traj}), automatic source identification and 
cross-matching algorithms based on triangulation matching \citep[see also][]{pal2006} 
were not found to be effective\footnote{The triangle parameter space 
is highly biased in this case while triangulation matching algorithms works 
most efficiently when the triangles have a nearly uniform 
distribution in this space.}.
Therefore, we manually selected seven 
prominent\footnote{Bright, isolated 
stars which are close to the center of the stripes.} stars distributed 
nearly uniformly in the field. Namely, we picked four stars 
located in the northern stripe and three stars in the southern stripe. 
Initial pixel coordinates for these stars have been acquired using the 
\texttt{imexam} tool of IRAF\footnote{IRAF is distributed by the 
National Optical Astronomy Observatory, which is operated by the 
Association of Universities for Research in Astronomy (AURA) under 
a cooperative agreement with the National Science Foundation.} on 
adjacent frame pairs belonging to events where the pointing of 
{\it Kepler} was changed abruptly. We found that this kind of manual 
analysis of $39$ images were sufficient in order to bootstrap the 
astrometry and for the cross-identification procedure for all of the $3789$ 
individual images. 

After the cross-identification and the derivation of the differential
transformations were performed, we used these transformations to register
the images to the same system. Then, we stacked every tenth of the images
after correcting of the background variations in order to obtain a master
frame. This master image was created using a median averaging  
to exclude the effects of the cosmic hits, as well as to safely 
rule out the contribution of other moving objects\footnote{See 
http://szofi.net/apal/astro/k2/tno/ for animations.}.

\begin{deluxetable}{lcrr}
\tabletypesize{\scriptsize}
\tablecolumns{8}
\tablewidth{0pc}
\tablecaption{Photometric data of \jjs{} and \gv{}\label{table:phot}}
\tablehead{Object & \colhead{Time (JD)} & \colhead{Magnitude\ensuremath{^{\rm a}}} & \colhead{Error}}
\startdata
\jjs{}	&	2456893.044947 & 20.288 & 0.078 \\
--	&	2456893.208405 & 20.686 & 0.111 \\
--	&	2456893.228837 & 20.722 & 0.105 \\
\hline
\gv{}	&	2456839.65 & 23.680 & 0.606 \\
--	&	2456839.75 & 22.862 & 0.402 \\
--	&	2456839.85 & 22.957 & 0.403 
\enddata
\tablecomments{Table \ref{table:phot} is published in its entirety in the
electronic edition of the {\it Astrophysical Journal Letters}.  A portion is
shown here for guidance regarding its form and content.}
\tablenotetext{a}{Magnitudes shown here are transformed to USNO-B1.0 $R$
system, see text for further details.}
\end{deluxetable}

Photometry of \jjs{} was then performed using differential images 
(difference between the individual images and the aforementioned master image)
where the aperture centroid coordinates were based on Kepler-centric apparent
positions retrieved from the NASA/Horizons 
service\footnote{http://ssd.jpl.nasa.gov/horizons.cgi}. 
The instrumental aperture was chosen to be relatively small due to the 
faintness of the object. Despite the usage of differential images, the 
residual structures were significant due to the undersampled property of 
the {\it Kepler} camera. Namely, the spline-based interpolation throughout 
the registration process always introduces undershootings in case of stellar 
profiles with small FWHMs yielding characteristic features on the residual 
images \citep[see e.g. Fig. 4 in][]{pal2009}. Photometric uncertainties have 
been estimated considering photon noise and measured
background scatter in the aperture annuli. For the photon noise estimation,
we used the gain values reported in the original FITS header, namely 
$113.3\,{\rm e^-/ADU}$. As we will discuss later on 
(Sec.~\ref{sec:analysis}), this estimation was found to be 
reliable. 

In the case of \gv{}, we followed a similar data flow for image reduction,
with the exception that the frame registration was much easier due to the
smaller size and the fewer number of stars (Fig.~\ref{fig:gvimage}). Namely, 
the number of unambiguously identified background stars was three and the 
mean of the centroid shifts (in both directions) was used to obtain the 
offset values for the image registration. In addition, we binned the images
into intervals of $0.1{\rm d}$ in order to increase the signal in the frames.
Hence, each frame on which photometry is performed were derived from $\sim 5$
individual downlinked measurements. In all other aspects, the 
photometry was performed identically for the two objects.

For both objects, photometric magnitudes have been transformed into USNO-B1.0 
$R$ system \citep{monet2003}. In the case of \jjs{}, we found that the 
unbiased residual of the the photometric transformation between USNO and
{\it Kepler} unfiltered magnitudes was $0.05\,{\rm mag}$,
while for \gv{} it was found to be $0.09\,{\rm mag}$. Stars used for both
objects have magnitudes spanning nearly homogeneously 
between $R\approx 15$ and $R\approx 19$. Photometric data are displayed
in Table~\ref{table:phot}.

\begin{figure}
\centering\includegraphics[width=5.5cm,angle=270]{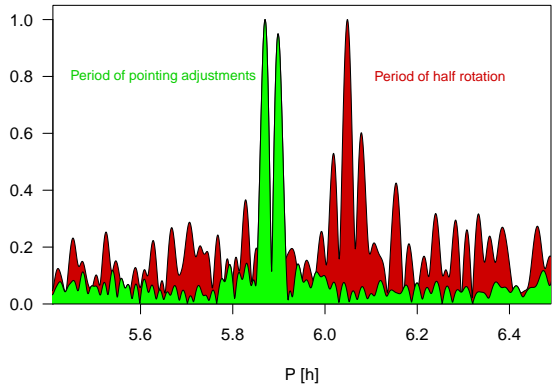}
\caption{Fourier transform of the K2 light variation of 
the \jj{} normalized to 1 (red). The frequency axis was transformed to 
the time (period) dimension. In order to demonstrate that thanks to the 
motion of the TNO, the pointing adjustments were averaged out from the 
light variation, we plot the Fourier transform of the $x$
coordinate of the centroid. The variations from the TNO rotation and 
the spacecraft pointing adjustment are clearly separated.}
\label{fig:fourier}
\end{figure}

\section{Light curve analysis}
\label{sec:analysis}

\begin{figure*}
\centering\includegraphics[width=5.8cm,angle=270]{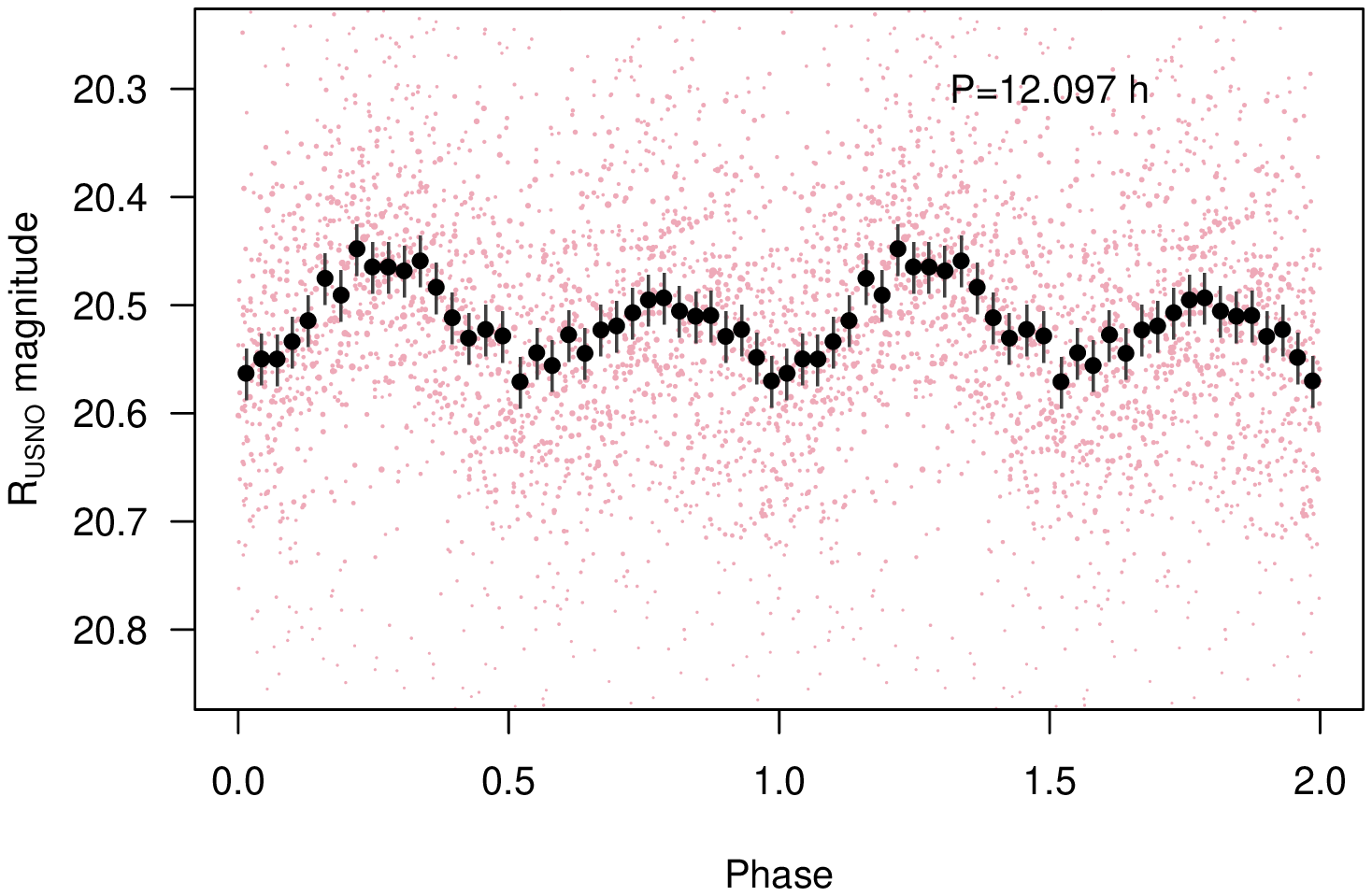}
\centering\includegraphics[width=5.8cm,angle=270]{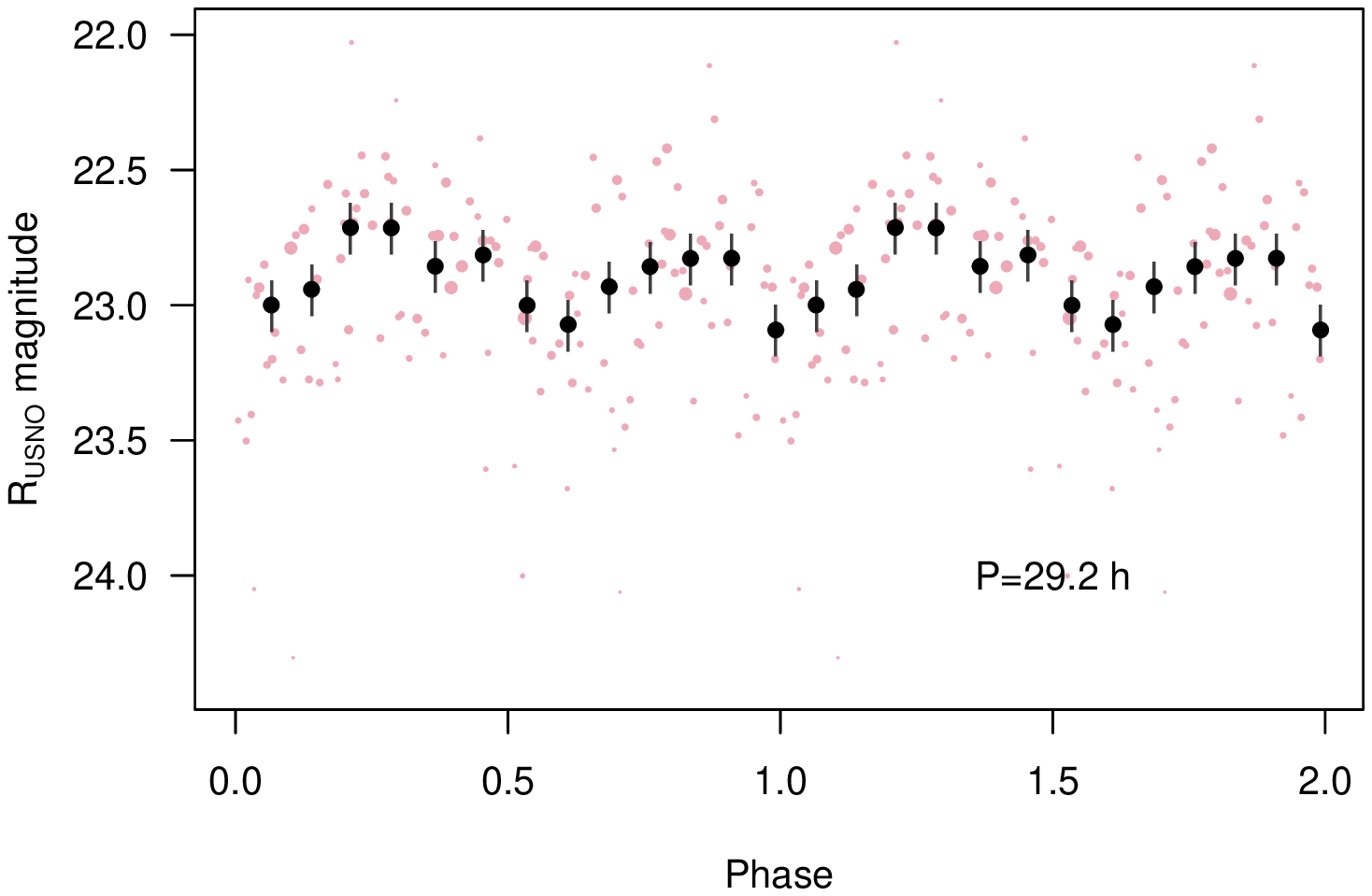}
\caption{Left: phased light curve of 
\jj{} (red points). The bold symbols with the error bars are binned 
values. Right: K2 light curve of \gv{}.}
\label{fig:lc}
\end{figure*}

The data produced from the baseline photometry consisted of
time instance, magnitude and magnitude errors. 2224 data points 
belonged to \jjs, and 129 points to \gv, respectively. We tested that 
the photometric error estimates by the reduction pipeline are quite 
realistic. The mean determined photometric error of the entire dataset 
is $0.163$ magnitudes, while the mean of the standard deviation of data 
points in the boxcar is $0.127$ magnitudes. These values are comparable 
to each other within 30\%{} accuracy. For later processing, the weights 
of the photometric data were set to be proportional to $1/\sigma_i^2$, 
where $\sigma_i$ is the formal photometric error of each data points.

We also observed that this approach overweights the outliers on the 
bright side, and may introduce a distortion in the shape of the light 
curve. This is because error sources, such as residuals after image 
subtraction, scatter toward both directions (too faint and too bright), 
but the mean brightness is biased here, because the formal errors of 
bright-looking objects are much smaller, and they are given larger 
weights. Therefore at the bright wing of data distribution, the weights 
were downscaled by a function of the shape 

\begin{equation}
\arctan\left[8\,e^{-d^2/\mu^2}\right]+0.3
\end{equation} 

where $d=m-m_0$, the deviation of the measured $m$ magnitudes from the 
mean value. We selected $\mu\approx4$ times the peak of the Fourier 
spectrum at the rotational period ($0.15$ for \jjs{} and $0.5$ for \gv{}, 
respectively). Here the peak of the Gaussian curve is damped by 
the $\arctan(\cdot)$ function acting on it, leading to a flat 
plateau around the center of the shape. We are convinced that this 
rescaling leads to more realistic weights than 
the crude estimates making use of the measured flux values. However, 
in order to avoid biasing the results in any way, we compared the results 
with and without weighting, and we experienced that they were 
practically identical -- so the modified weights supported 
merely the better visualization.

The rotation periods were found by a Fourier analysis. The Fourier 
transform of the light curves were calculated by the {\tt Period04} 
code \citep{lb05}. To exclude the light variation introduced by periodic 
systematics, we also calculated the periodograms of the pointing position 
of the telescope, the actual and the measured subpixel positions of the 
TNOs in each frames. We did not find periodic signals resulting from the 
sampling effects (therefore subpixel positions), but we identified the 
peak due to the pointing corrections with $4.05\,{\rm cycle}/{\rm d}$ 
frequency (Fig.~\ref{fig:fourier}). This is close to the frequency peak 
belonging to a half rotation of \jjs{}, but they are still well 
separated, proving that the periodograms of the TNOs reflect the rotation.

\section{Results}
\label{sec:results}

\subsection{\jj}
\label{sec:jj}

\jj{} is an outer Kuiper belt object, likely in $2:1$ resonance with 
Neptune \citep{mpc60235}. Based on its absolute magnitude at 
discovery ($H_{\rm V}=3.9$) and the cluster average geometric 
albedo of $p_{\rm V}=0.13_{-0.07}^{+0.09}$ for objects in the 
outer Kuiper belt \citep{lacerda2014}, its likely effective diameter 
is $D_{\rm eff}=610_{-140}^{+170}\,{\rm km}$. 
\jjs{} was also detected by the Southern Sky and Galactic Plane Survey 
for Bright Kuiper Belt Objects \citep{sheppard2011}, and an R-band 
absolute magnitude of $H_{\rm R}=3.2$ was associated to the object.

\cite{bs2013} detected a light curve of \jjs{}, with a peak-to-peak 
amplitude of $0.13\pm0.02$ in the Sloan {\it r'} band. The most likely light 
curve (rotation) period they found was $6.04\,{\rm h}$, or similarly 
likely its double period, $12.09\,{\rm h}$. However, other 
periods ($4.83/9.66\,{\rm h}$) were also possible. The light curve was 
clearly non-sinusoidal. The absolute magnitude in this band was found to 
be $H_{r'}=4.17\pm0.20$.

The period analysis confirmed the most likely solution of \cite{bs2013}, 
namely $6.048\pm0.018\,{\rm h}$. This is clearly the half of the rotation period, 
confirmed by the asymmetry of the phased light curve with $P=12.097\pm0.036\,{\rm h}$ 
period (left panel of Fig.~\ref{fig:lc}). Weighted means of the magnitude 
and corresponding phases were calculated in a boxcar of 68 data points. 
The standard deviation and the number of points in each bins resulted in 
an expected standard deviation (error bars) of $\approx0.015$ magnitudes 
for the binned points. The smoothed light curve shows two humps with 
equivalent minima and slightly different peaks. The full amplitude is 
$0.100\pm0.005$ magnitude, the difference between the minima is $0.02$ 
magnitudes. The asymmetry and the non-sinusoidal shape of the light 
curve reflects the uneven surface structures. These features are quite 
common for Solar System objects, e.g. a very similar light curve shape 
is observed for the main belt asteroid (52)~Europa at certain observing 
circumstances \citep[e.g.][]{michalowski2004}. We can also conclude that 
the presented K2 measurement and the previous photometry of \jjs{} 
in \cite{bs2013} shows similar characteristics since in the case of TNOs,
the phase and aspect angles varies rather slowly on the timescale of
these observations.

\subsection{\gv}
\label{sec:gv}

\gv{} is a dynamically cold classical Kuiper belt object. 
A recent evaluation of 
Johnson R-band MPC data and a conversion using $V-R=0.59\pm0.15$ 
(average for classicals) resulted in a V-band absolute magnitude 
of $H_{V}=6.1\pm0.6$. \gv{} was observed, but not detected in any 
bands of the PACS camera on board the Herschel Space Observatory, 
in the framework of the "TNOs are Cool!" Open Time Key Program 
\citep{mueller2009}. Using the upper limits set by the non-detections, 
a modelling of the thermal emission of the target resulted in an upper 
limit of its size ($D\lesssim180\,{\rm km}$), and a lower limit on its 
geometric albedo \citep[$0.19\lesssim p_V$, see][]{vilenius2012,vilenius2014}. 
No light curve information is available for this target.

The first light curve of this object is shown in the right panel of 
Fig.~\ref{fig:lc}, phased with the most likely rotation period of 
$29.2\pm1.1\,{\rm h}$ and showing a double peak. Since we had $129$ photometric 
points from K2 in this case, they were binned into $13$ points during one 
rotation, and the error bar of the binned points resulted to be 
$0.06$ magnitudes. The full amplitude of the light variation is $0.35$ 
magnitudes, which can be a sign of a more exposed asphericity than for 
\jjs{}. 
The most interesting feature of this TNO is its
long rotation period, as most TNOs and Centaurs have rotation periods shorter
then a day \citep{duffard2009,bs2013}. 

Longer periods are expected from synchronously locked, tidally evolved binary 
systems \citep[see Pluto/Charon or Sila/Nunam,][]{grundy2012}. It can 
explain the slow rotation of \wg{} \citep{rabinowitz2013} and can be 
plausible for \gv{} either. Moreover, observational statistics can explain 
binarity more likely for dynamically cold TNOs \citep{noll2008}. In addition, 
statistics seem to prefer close orbits \citep[Fig. 4 in][]{noll2008} which
can also yield mutual events and hence strong light curve variations.

\section{Conclusions}
\label{sec:conclusions}

We clearly detected the rotational signal of two TNOs with {\it Kepler}
during the K2 extended ecliptic survey mission. 
Since these objects are moving through the field-of-view, photometric
data reduction requires special care in order to retrieve accurate 
flux information. Although the image processing needs complex steps,
our results clearly show that the moving nature of these targets 
are also advantageous. Namely, differential photometric techniques 
strenuously reduce the effects of background sources, even in the 
crowdness of {\it Kepler} fields and in the case of lower imaging resolution
and undersampled stellar profiles. In addition, the employment of
moving apertures removes the frequency aliases caused by the 
pointing corrections of the spacecraft in every $\sim 6$ hours. 

All in all, we successfully demonstrated that it is feasible to observe faint 
TNOs around their stationary points with {\it Kepler}/K2 and this mission 
provides a unique way to monitor them continuously for many weeks. 
The scientific benefit of observing TNOs is the opportunity
to provide an unbiased sample of rotational light curves -- and hence
data about shape, albedo and surface characteristics. This kind of information
aids us to understand the nature and evolution of the outer Solar System.

Future space missions, like TESS and PLATO are not well suited to this 
kind of observations, therefore we encourage to include the brightest 
TNOs in each observing campaign to exploit this unique capability of 
the K2 Mission.

\vspace*{2mm}

\begin{acknowledgements}
We thank the hospitality of the Veszpr\'em Regional Centre of the 
Hungarian Academy of Sciences (MTA VEAB) where most of our work was carried out. 
We also thank the comments of the anonymous referee.
This project has been supported by the Lend\"ulet-2009 and 
LP2012-31 Young Researchers Program,
the Hungarian OTKA grants K-83790, K-109276 and K-104607 and by City of 
Szombathely under agreement no. S-11-1027. 
The research leading to these results has received funding from the 
European Community's Seventh Framework Programme (FP7/2007-2013) under 
grant agreements no. 269194 (IRSES/ASK), no. 312844 (SPACEINN), and the ESA 
PECS Contract Nos. 4000110889/14/NL/NDe and 4000109997/13/NL/KML. 
Gy.~M.~Sz. and Cs.~K. was supported by the J\'anos Bolyai Research 
Scholarship. Funding for the K2 spacecraft is provided by the NASA Science 
Mission directorate. 
The authors acknowledge the Kepler team for the extra efforts to 
allocate special pixel masks to track moving targets. 
All of the data presented in this paper were obtained from the 
Mikulski Archive for Space Telescopes (MAST). 
STScI is operated by the Association of Universities for Research in 
Astronomy, Inc., under NASA contract NAS5-26555. Support for MAST for 
non-HST data is provided by the NASA Office of Space Science via 
grant NNX13AC07G and by other grants and contracts.
\end{acknowledgements}


{}

\end{document}